\documentclass{elsart}   

\def\Journal#1#2#3#4{{#1} {\bf #2}, #3 (#4)}

\def\NPB{{\em Nucl. Phys.} B}

\def\PRL{\em Phys. Rev. Lett.}
\def\PRD{{\em Phys. Rev.} D}

\usepackage{graphicx} 
\usepackage{epsfig}  
\usepackage{epsf,amsfonts,amssymb,amsbsy}
\usepackage{wrapfig}
\RequirePackage{subfigure}
   
\begin{document}   
\begin{frontmatter}   
\title { Expected Accuracy in a Measurement of the CKM Angle $\alpha$ Using 
a Dalitz Plot Analysis of $\boldmath B^0 \to \rho \pi$ Decays in the 
BTeV Project}
\author[IHEP]{K.E.~Shestermanov\thanksref{remark}},   
\thanks[remark]{untimely deceased}
\author[IHEP]{A.N~Vasiliev\thanksref{addr}},   
\thanks[addr]{corresponding author, email: Alexander.Vasiliev@ihep.ru}   
\author[FNAL]{J.~Butler},   
\author[IHEP]{A.A.~Derevschikov},   
\author[FNAL]{P.Kasper},  
\author[IHEP]{V.V.~Kiselev},   
\author[IHEP]{V.I.~Kravtsov},
\author[UMN]{Y.~Kubota},  
\author[FNAL]{R.Kutschke}, 
\author[IHEP]{Y.A.~Matulenko},   
\author[IHEP]{N.G.~Minaev},   
\author[IHEP]{V.V.~Mochalov},   
\author[IHEP]{D.A.~Morozov},   
\author[IHEP]{L.V.~Nogach},
\author[INFN-Milano]{M.~Rovere},   
\author[IHEP]{A.V.~Ryazantsev},   
\author[IHEP]{P.A.~Semenov},   
\author[SYR]{S.~Stone},   
\author[IHEP]{A.V.~Uzunian},   
\author[FNAL]{J.~Yarba}   
\collab{BTeV Electromagnetic Calorimeter Group}   
\date{\today}   
   
\address[IHEP]{Institute for High Energy Physics, Protvino 142281, Russia}   
\address[FNAL]{Fermilab, Batavia, IL 60510, U.S.A.}   
\address[UMN]{University of Minnesota, Minneapolis, MN 55455, U.S.A.}   
\address[SYR]{Syracuse University, Syracuse, NY 13244-1130, U.S.A.} 
\address[INFN-Milano]{Istituto Nazionale di Fisica Nucleare - Sezione di Milano, Milano I-20133, Italy}  
   
\begin{abstract}   
A precise measurement of the angle $\alpha$ in the CKM
triangle is very important for a complete test of Standard Model.
A theoretically clean method to extract $\alpha$ is provided by 
B$^0 \to \rho \pi$ decays. Monte Carlo simulations to obtain the BTeV 
reconstruction efficiency and to estimate the signal to background ratio 
for these decays were performed. Finally the time-dependent Dalitz 
plot analysis, using the isospin amplitude formalism for tree and 
penguin contributions, was carried out. It was shown that in 
one year of data taking BTeV could achieve an accuracy on $\alpha$ 
better than 5$^{\circ}$.

\end{abstract}    
\end{frontmatter}   
   
\newpage   


\section{Introduction}

The Standard Model (SM)~\cite{langacker} which incorporates the quark mixing
Cabbibo-Kabayashi-Maskawa (CKM) mechanism~\cite{ckm} has been increasingly
successful, supported with many precise experimental results. This strongly
indicates that, at low energies, the SM is the effective description of
Nature. However, there are reasons to believe that there exists physics 
beyond the SM. For example, from the astrophysical point of view, it is
a serious problem that the matter-antimatter asymmetry in the Universe 
can not be explained solely from the CP violation in the SM, which originates
from quark flavor mixing.
This observation, together with others, leads one to believe that there is 
a new physics,
most likely, at the TeV energy scale. One of the critical measurements
to test the SM or to obtain strong indications of new physics are
precisie measurements of the angles in the Unitary Triangles (UT), which 
are of non-zero area  if CP violation exists.

A unique program that would have allowed one to challenge the SM explanation 
of CP violation, 
mixing and rare decays in the $b$ and $c$ quark system was proposed by the BTeV 
project \cite{prop} at the Tevatron at Fermilab. The design of BTeV exceled in 
several crucial areas including: triggering on decays with purely hadronic final 
states, charged particle identification, excellent electromagnetic calorimetry 
and excellent proper time resolution. 
Exploiting the large number of $b$'s and $c$'s produced at the Tevatron collider, 
the experiment would have provided precise measurements of SM parameters 
and an exhaustive search for physics beyond the SM. 
The complete physics objectives of BTeV included measuring the CP violating angles
$\alpha$, $\beta$, and $\gamma$ of the UT. In particular, the measurement of $\alpha$ is 
difficult due to small overall rates and because the gluonic penguin rates are 
of the same order as the tree rates, causing well known difficulties in extracting 
the weak phase angle.
Quinn and Snyder~\cite{Snyder:1993mx} have suggested a theoretically clean 
method to extract $\alpha$ from decays of the type $B^0 \rightarrow \rho\pi$. 
The final state of these decays is not a CP eigenstate, which results in the
need of a Dalitz plot analysis.
We focus on the measurement of $\alpha$, via collecting a large sample 
of $B^0 \rightarrow (\rho\pi)^0$ decays. Direct measurements from 
the B-factories demonstrate that the average $hh$ decays (where $h = (\rho, \pi)$)
is known to a precision of $O(10^o)$~\cite{bevan}, with the use of isospin. In this paper
we demonstrate that BTeV could have done the measurement with a much better
precision. 

The paper is organized as follows. Section 2 provides brief introduction of
the CKM matrix. In section 3 we give a general overview of the BTeV project.
In sections 4 and 5, respectively, we report on the expected reconstruction 
efficiencies and the Signal/Background ratio of the $B^0 \rightarrow (\rho\pi)^0$ 
decays in BTeV. 
Section 6 covers the phenomenological formalism of $B^0 \rightarrow (\rho\pi)^0$
decays. In section 7 we describe results of the time-dependent Dalitz plot 
analysis of the simulated  $B^0 \rightarrow (\rho\pi)^0$ decays in BTeV.
 
\section{ The CKM Matrix and the Angle $\boldmath\alpha$}

In the SM there are three generations of leptons
and quarks.
The physical point-like particles that have both strong and 
electroweak interactions, the quark, are mixtures of weak eigenstates, 
described by a 3x3 unitary matrix, called the Cabibbo-Kobayashi-Maskawa~(CKM) 
matrix \cite{ckm},
\begin{equation}
\left(\begin{array}{c}d'\\s'\\b'\\\end{array} \right) =
\left(\begin{array}{ccc} 
V_{ud} &  V_{us} & V_{ub} \\
V_{cd} &  V_{cs} & V_{cb} \\
V_{td} &  V_{ts} & V_{tb}  \end{array}\right)
\left(\begin{array}{c}d\\s\\b\\\end{array}\right)
\end{equation}
The unprimed states are the mass eigenstates, while the primed states denote
the weak eigenstates. The $V_{ij}$'s are complex numbers that can be
expressed in terms of four independent real quantities.
These numbers are  fundamental constants of Nature that 
need to be determined from experiment.
In the Wolfenstein 
approximation the matrix is written as \cite{wolf}
\begin{equation}
V_{CKM} = \left(\begin{array}{ccc} 
1-\lambda^2/2 &  \lambda & A\lambda^3(\rho-i\eta(1-\lambda^2/2)) \\
-\lambda &  1-\lambda^2/2-i\eta A^2\lambda^4 & A\lambda^2(1+i\eta\lambda^2) \\
A\lambda^3(1-\rho-i\eta) &  -A\lambda^2& 1  
\end{array}\right)
\end{equation}
where ($\lambda$, $A$, $\rho$, $\eta$) are four mixing parameters with
$\lambda = |V_{us}| \approx 0.22$, $A \approx 0.8$ (measured using 
semileptonic $s$ and $b$ decays \cite{virgin}), and $\eta$ represents 
the CP violating phase.
This expression is accurate to order $\lambda^3$ in the real part and
$\lambda^5$ in the imaginary part. It is necessary to express the matrix
to this order to have a complete formulation of the physics we wish to pursue.

\begin{figure}[htb]
\centerline{\epsfig{figure=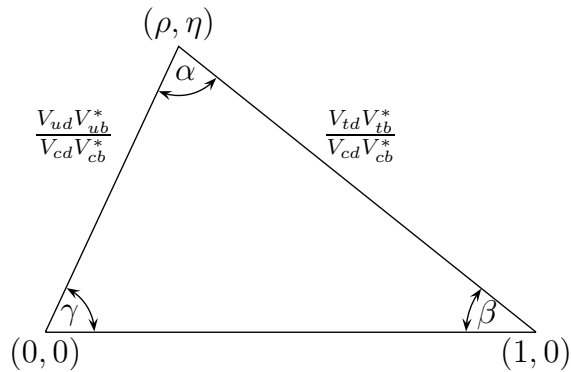,height=2.0in}}
\caption{\label{ckm_tri}Unitary Triangle.}
\end{figure}

The unitarity of the CKM matrix
leads to various relations among the matrix elements 
\begin{equation}
V_{ud}V_{us}^*+V_{cd}V_{cs}^*+V_{td}V_{ts}^*=0
\end{equation}
\begin{equation}
V_{us}V_{ub}^*+V_{cs}V_{cb}^*+V_{ts}V_{tb}^*=0
\end{equation}
\begin{equation}
V_{ub}V_{ud}^*+V_{cb}V_{cd}^*+V_{tb}V_{td}^*=0
\label{eq:UT}
\end{equation}
that can be geometrically represented in the complex plane as triangles. 
These are ``unitary triangles'', though the term ``unitary triangle'' 
is usually reserved only for the {\bf bd} triangle in (\ref{eq:UT}) where 
the angles are all thought to be relatively large.
This CKM triangle is depicted in Figure~\ref{ckm_tri}. 
It shows the angles
$\alpha,~\beta$, and $\gamma$. These angles are defined as
\begin{eqnarray}
\alpha=arg\left[-{V_{td}V^*_{tb} \over V_{ud}V^*_{ub}}\right], ~~~
\beta=arg\left[-{V_{cd}V^*_{cb} \over V_{td}V^*_{tb}}\right], ~~~
\gamma=arg\left[-{{V_{ud}V^*_{ub}}\over {V_{cd}V^*_{cb}}}\right]
\end{eqnarray}
and can be determined by measuring CP violation in $B$ decays. 



They can roughly be divided in two classes :
\begin{itemize}
\item{ decays that are expected to have relatively small direct CP violation
and hence are particularly interesting for extracting CKM parameters from
interference of decays with and without mixing.}
\item{ decays in which direct CP violation could be significant and therefore
that can not be cleanly interpreted in terms of CKM phases.}
\end{itemize}
$B$ decays used to extract $\beta$ belong to the first group, whereas decays
that have been considered to measure $\alpha$ belong to the second one.

The primary source for measurements of sin($2\beta$) are the decays 
of the type $b \to c \overline c s$.
The most statistically significant measurements of CP violation in 
the $B$ system were made by BABAR~\cite{babar} and BELLE~\cite{belle},
resulting in the average value 
of sin($2\beta$)=0.725$\pm$0.037~\cite{hfag}.


Measuring $\alpha$ is more difficult than measuring $\beta$ in several respects. 
First of all, the decay amplitudes are modulated by $V_{ub}$ rather than $V_{cb}$ 
making the overall rates small, of the order of $10^{-5}$ to $10^{-6}$. Secondly, 
the gluonic penguin rates are of the same order as the trees causing large theoretical 
uncertainties in cleanly extracting $\alpha$ from asymmetry measurements, alone.

The decay $B^0 \rightarrow \pi^+\pi^-$ has been proposed as a way to measure 
sin($2\alpha$). However, the penguin pollution is quite large and cannot be 
ignored. Gronau and London~\cite{GRL} have shown that an isospin analysis 
using the additional decays $B^- \rightarrow \pi^-\pi^0$ and $B^0 \rightarrow \pi^0\pi^0$
can be used to extract $\alpha$~\cite{pipimodel}, but the $\pi^0\pi^0$ final state
is extremely difficult to detect in any existing or proposed experiment.
$B \to \pi \pi$ has been seen but there is no $B$ decay vertex information,
therefore there is no way to perform a time-dependent CP violation measurement.
If fact, the data that does exist has been used to limit the Penguin
contribution to these decays, but the limit is not very restrictive. 
Lipkin, Nir, Quinn and Snyder~\cite{LNQS} have extended the analysis 
in~\cite{GRL} to include other decays, among them $B \rightarrow \rho\pi$. 
Snyder and Quinn~\cite{Snyder:1993mx} subsequently extended that work and 
proposed not only an isospin analysis, but a full, time dependent, Dalitz 
plot study of $B \rightarrow \rho\pi$ decay to measure $\alpha$.

\begin{figure}[htb]
\centerline{\epsfig{figure=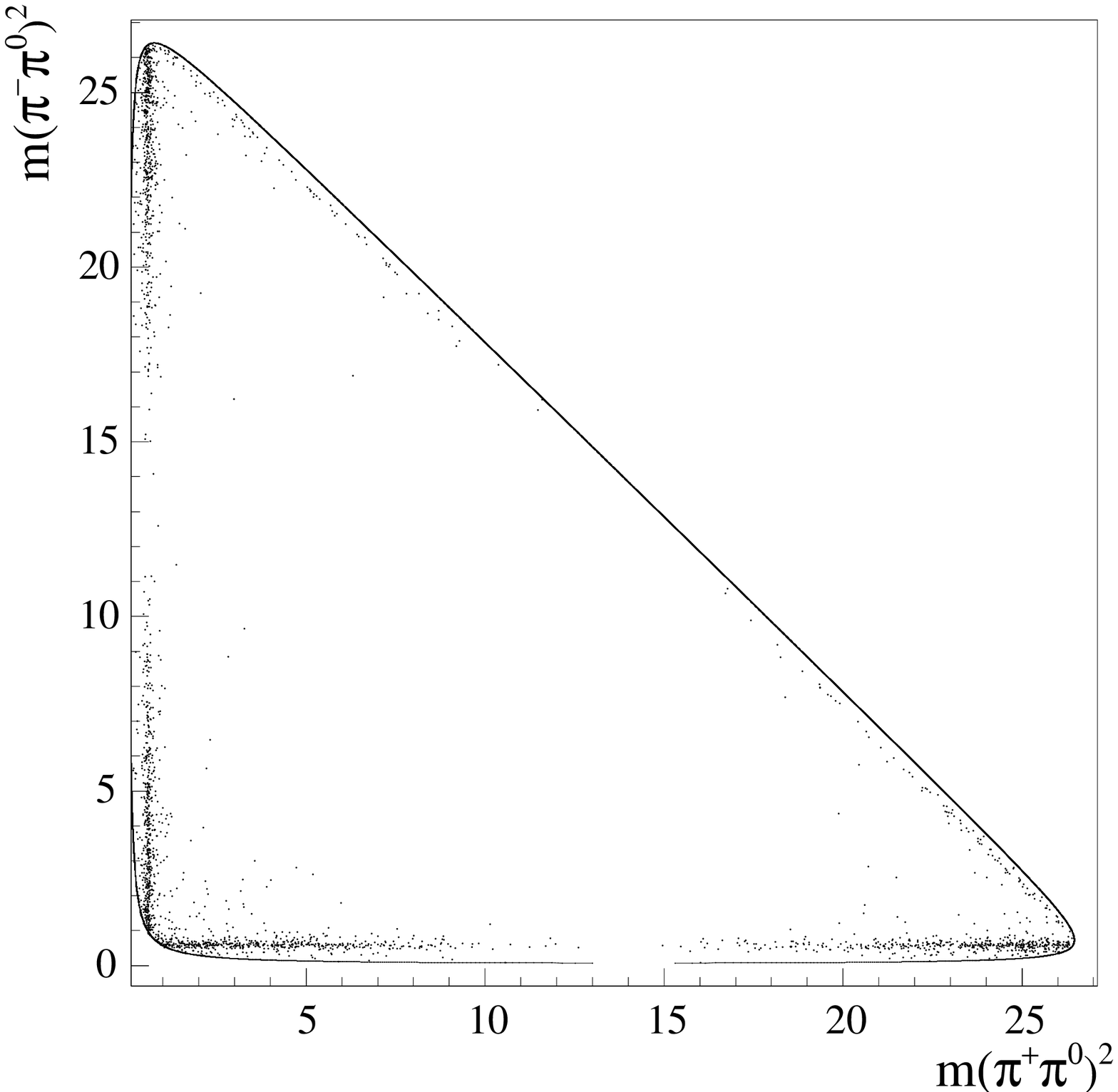,height=2.8in}}
\caption{\label{fig:dalitz} The Dalitz plot for $B^o\to\rho\pi\to\pi^+\pi^-\pi^o$.}
\end{figure}

A sample Dalitz plot is shown in Fig.~\ref{fig:dalitz}. 
A striking feature of this Dalitz plot is that the events are concentrated
close to the kinematic boundary, especially in the corners.
This kind of distribution is good for maximizing the interference, which 
helps minimize the errors. Furthermore, little information is lost 
by excluding the Dalitz plot interior, a good way to reduce backgrounds.

Snyder and Quinn have performed an idealized Dalitz plot analysis
that uses 1000 or 2000 flavor tagged background free events. 
Trials using 1000 events usually yield good results for $\alpha$, 
but sometimes do not resolve the ambiguity. With the 2000 event 
samples, however,the ambiguities disappear.

Recently BABAR has made an important step to improve the constraints
on $\alpha$, via studying $B^0$({\it \=B}$^0$)$\rightarrow \rho^+\rho^-$ 
decays~\cite{babar_0503049}.  
Using the isospin analysis they determined that the solution compatible with 
the Standard Model is $\alpha = (100 \pm 13)^o$.
The estimate is based on 232 millions $\Upsilon (4S) \to B \overline B$ decays.
This mode has potential show stoppers to improving errors on $\alpha$.
The analysis assumed a 100\% longitudinal polarization of the 
$B^0 \rightarrow \rho^+\rho^-$; if this is not true, an angular analysis 
is needed and requires a lot more data. 
However, BABAR measured the longitudinal polarization fraction 
$f_L = 0.978 \pm 0.014 (stat) ^{+0.021}_{-0.029}(syst)$~\cite{babar_0503049}
which is consistent with one.

Regardless of that the $\rho\pi$ system remains theoretically the cleanest 
way to extract $\alpha$. Recently BABAR has performed first full time dependent
Dalitz plot analysis~\cite{babar_0503049} and extracted $\alpha = (113^{+27}_{-17} \pm 6)^o$. 
In the following sections 
we will show how, with the BTeV detector, one would have significantly improved 
sensitivity on $\alpha$ using the full time-dependent Dalitz plot analysis
of the $B \rightarrow \rho\pi$ decays.

\section{The BTeV Concept} 
 
BTeV was designed as a second generation experiment to study CP violation 
in $B$ decays. It would have made possible to carry out practically all 
measurements of CP violation and decays of the $B$ hadrons accessible 
the asymmetric $B$-factories and at CDF and D0 running at Tevatron 
and it could have done those measurements at a much higher precisions.
The detector design is ideally suited to study $B$~decays containing 
neutral particles, especially the modes of interest here, 
$B^0 \to (\rho \pi)^0$.
Towards the end of the decade, LHCb will go into operation with similar 
capabilities for all-charged states, but will not have a high-quality
calorimeter or a trigger which is efficient for all $B$~decay modes. 

The studies presented in~\cite{CDFcentral,D0central,MNR,beamdrag,artuso} 
indicate that 
the forward direction at the Tevatron presents a number of striking advantages. 
First of all, there is a large cross-section for the production of correlated 
$b\bar{b}$ pairs. Secondly, the $B$ hadrons
that are formed have relatively large momenta, on average 30 GeV/c, and,
therefore, their decay products do not suffer much from multiple 
Coulomb scattering. This would allow BTeV to make precision measurements 
of the spatial origins of particles and as result BTeV would be able to determine 
if they arose from $B$ hadrons that traveled on the order of several mm prior 
to their decays. Furthermore the geometry 
was very natural for certain aspects of detector technology that significantly 
enhance the physics performance.
For these reasons, the BTeV collaboration designed a detector with ``forward 
coverage.''

The physics case for BTeV involves reconstructing a variety of different 
decay modes of the $B$, $B_{s}$, and other $b$ hadrons and, in many cases, 
following their time evolution and tagging the flavor of the parent $B$ at 
production and at the moment of decay. These decay modes may involve charged 
hadrons, charged leptons, photons (prompt or from $\pi^{o}$'s), and tertiary
vertices from the $b\rightarrow c$ decay chain. 
The product branching fractions of many decay modes of interest, 
including any tertiary decays, are quite small, typically $10^{-5}$ to $10^{-7}$. 
This, together with the large background of minimum bias events, demanded that
BTeV be able to reconstruct multibody final states, with a good resolution
in invariant mass, and to handle very high data rates. In order to carry out
the physics program, the detector must have the ability to separate decay 
vertices from the primary interaction vertex and to reconstruct secondary 
$B$ vertices and daughter charm vertices. This requires a precision vertex 
detector. It must also be able to measure the time evolution of decays for 
time-dependent asymmetry studies. The most demanding requirement is to be able 
to follow the very rapid oscillations of the $B_{s}$ meson in order to study 
mixing and CP violation. The detector must have the ability to distinguish pions, 
kaons, and protons from each other to reduce confusion among similar decays such 
as $B \to \pi \pi$ and $B \to K \pi$ so that decays of interest will not be
contaminated by other decays, causing the resulting measurements to be diluted.
Many key decay modes 
have $\pi^{o}$'s, $\gamma$'s, 
or particles that decay into them, such as $\rho$'s or $\eta$'s. Leptons,
muons and electrons (positrons), appear in many key final states so good
lepton identification is also required. Finally, many of the detector
properties which are needed to isolate and reconstruct signals are
also needed to perform ``flavor tagging.''

\begin{figure}[htb]
\centerline{\epsfig{file=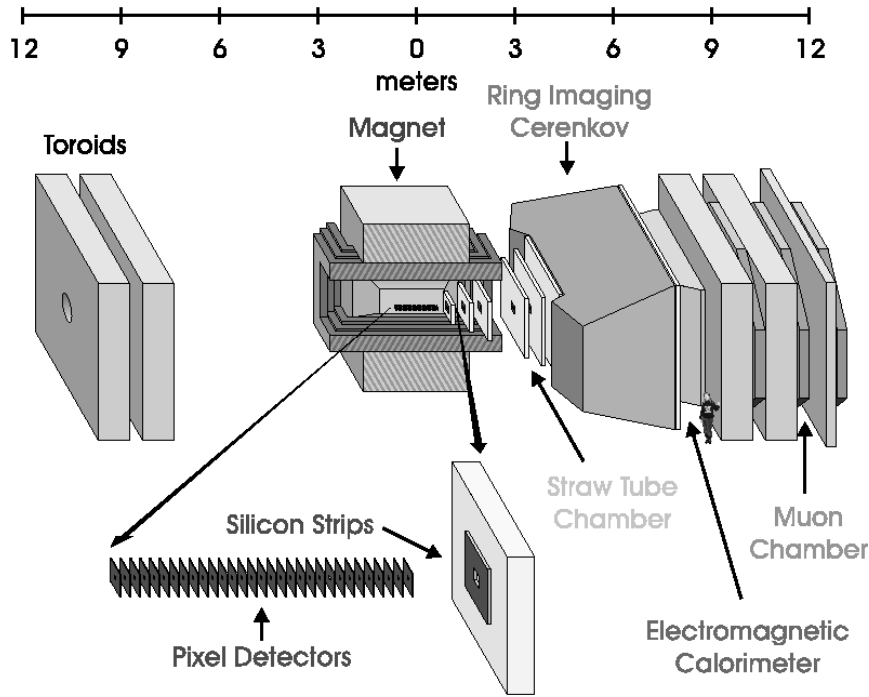,width=4.5in}}
\caption{Layout of the BTeV  Detector 
\label{intro_descoped_fig_1}}
\end{figure}

The BTeV detector is shown schematically in Fig.~\ref{intro_descoped_fig_1}. 
The covered angular region is from approximately 10 mr to 300 mr with respect 
to the anti-proton beam. 
When a $B$ decay of interest is contained within the acceptance of the
detector, there is a high probability that the decay products of the
co-produced $\bar{B}$ will also be within the acceptance of the detector.
Furthermore, since the charged $B$ decay products are not degraded by 
multiple scattering in the detector material, which allows accurate 
determinations of $B$ decay vertices.

The key design  features of BTeV include:
\begin{itemize}
\item A dipole centered at the interaction region placing a magnetic field 
on the vertex detector, allowing the use of momentum determination in the 
trigger. There are two open ends of the magnet. One open end allows particles 
to flow into the instrumented ``arm." The field is used by the tracking 
system to provide precise momentum determinations of all of the charged 
particles.
\item A precision vertex detector based on planar pixel arrays. 
The outputs are used in the trigger processor to find detached heavy
quark decay vertices in the first level trigger.
They also provide precise and unambiguous three-dimensional 
space points to help reconstruct charged particles; 
\item Precision tracking using a combination of straw tubes and silicon 
microstrip detectors, inside the straws close to the beam line, where the
charged particle occupancies are the largest. This system, when coupled with 
the pixels, provides excellent momentum and mass resolution out to 300 mr;
\item Excellent charged particle identification using a Ring Imaging 
Cherenkov Detector (RICH). The RICH provides hadron identification from 
3-70 GeV and lepton identification from 3-20 GeV, out to the full aperture 
of 300 mr, which is crucial since the muon detector and calorimeter do not
cover the full solid angle covered by the RICH. The RICH has two independent 
systems sharing the same space. One has a gas (C$_4$F$_8$O) radiator and 
a Multianode Photomultiplier photon detector and the other has a liquid 
C$_5$F$_{12}$ radiator and a Phototube photon detector. Both systems work 
in the region of visible light;
\item A high quality PbWO$_4$ electromagnetic calorimeter with excellent 
energy resolution, position resolution and segmentation, covering up to 
200 mr, capable of reconstructing final states with single photons, 
$\pi^o$'s, $\eta$'s or $\eta'$'s, and identifying electrons;
\item Excellent identification of muons out to 200 mr using a dedicated 
detector consisting of a steel toroid instrumented with proportional
tubes. This system has the ability to both identify single muons above 
momenta of about 10 GeV/c and supply a dimuon trigger; 
\item A detached vertex trigger at Level 1 using the pixel detector information, 
which makes BTeV efficient for most final states, including purely hadronic modes. 
The trigger ignores low momentum tracks that have large multiple 
scattering and would thereby avoid creating false secondary vertices; and 
\item A very high speed and high throughput data acquisition system which
eliminates the need to tune the experiment to specific final states.
\end{itemize}


\section{Reconstruction Efficiencies for $B^0 \rightarrow (\rho\pi)^0$ 
decays in BTeV}

As stressed earlier, measuring the time dependent CP violating
effects in the decays  $B^0 \rightarrow (\rho \pi)^0 \rightarrow
\pi^+ \pi^- \pi^0$  provides a theoretically clean way to determine
the angle $\alpha$ of the Unitary Triangle, as shown by Snyder and 
Quinn \cite{Snyder:1993mx}.
We report on the expected performance of the BTeV detector for these decays, 
taking into account excellent reconstruction efficiency of the $\pi^0$'s 
that is made possible with the electromagnetic calorimeter based on PWO 
crystals.  

Excellent mass resolution in the $\pi^0$ reconstruction reduces 
the background significantly, particularly near 
the edges of the Dalitz plot where the $\rho\pi$ events lay.
In addition, good resolution in the proper decay time is crucial 
to determine the angle $\alpha$.

The reconstruction efficiencies for $B \rightarrow \rho \pi$ were studied 
using GEANT3-based simulation~\cite{geant3}. We generated two samples: 
250,000 $\rho^{\pm} \pi^{\mp}$ and 250,000 $\rho^0 \pi^0$. Both samples 
were generated with a mean of two Poisson distributed non-beauty interactions 
per beam crossing. This number of interactions
per beam crossing corresponds to running at the designed Tevatron luminosity 
$2 \times 10^{32}$cm$^{-2}$s$^{-1}$ and 132~ns bunch spacing.
It should be mentioned that we used two separate Monte Carlo samples only
to refine the selection procedure and determine the reconstruction
efficiency and signal-to-background ratios. For the Dalitz plot analysis
the interference between charged and neutral $\rho$-mesons was
simulated.  

The analysis relies on BTeV event reconstruction software packages, including 
track reconstruction based on the Kalman filter method, vertex reconstruction, 
and shower reconstruction.

With the use of the electromagnetic calorimeter, we would find many good
$\pi^0$ candidates. Photon candidates
are required to have minimum reconstructed energy of 1~GeV and pass 
a shower shape cut designed to reject hadronic showers.
We reduce the background
rate by insuring that the photon candidates are not too close to the
projection of any charged tracks to the calorimeter.

\begin{figure}
\centering
  \subfigure[]{\includegraphics[width=0.50\textwidth,height=75mm]{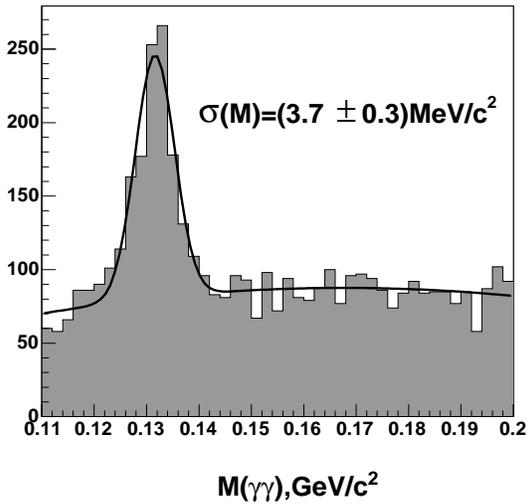}}
  \subfigure[]{\includegraphics[width=0.46\textwidth]{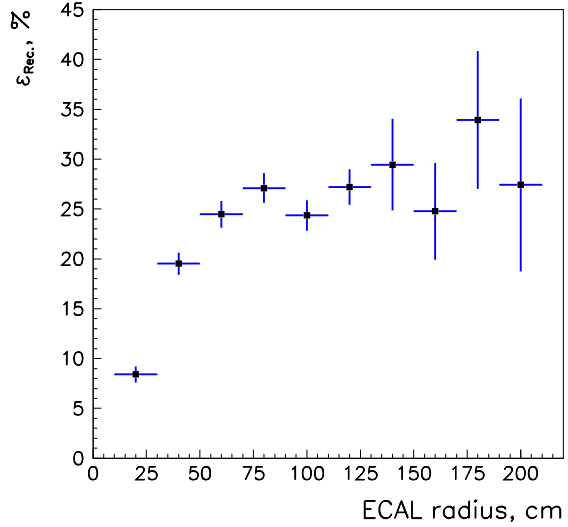}}
  \caption{The $\pi^0$ signal in the $\gamma \gamma$ spectrum from
  $B^0 \rightarrow \rho^+ \pi^-$ simulated events (a) and the efficiency of reconstructing
  the $\pi^0$ as a function of distance from the beam line (b). All events considered have
  at least one signal $B$ decay. The sample includes a mean
  of two Poisson distributed non-beauty interactions per beam crossing.}
  \label{fig:pi0_plot}
\end{figure}

Figure~\ref{fig:pi0_plot}a shows a $\gamma \gamma$ invariant mass distribution
of the $B \to \rho \pi$ events when the pairs have energy sum greater than 5~GeV
and the vector sum of transverse momenta greater than 0.75~GeV/$c^2$.
The $\pi^0$ signal is very clear; the $\pi^0$ mass resolution in this sample
is 3.7~MeV$/c^2$.

Candidate $\pi^0$'s are two-photon combinations with invariant
masses between 125 and 145~MeV$/c^2$. The $\pi^0$ reconstruction
efficiency depends on the distance from the beam line and is presented
in Figure~\ref{fig:pi0_plot}b; the $\pi^0$'s are taken from the $B^0
\rightarrow \rho^{\pm} \pi^{\mp}$ events; this simulation was 
run with a calorimeter larger than that proposed so we could view the
dependence on radius. The denominator contains all events in which 
the event passes the trigger, all charged tracks are reconstructed, 
and the combination of the two charged tracks passes some vertexing 
and detachment cuts. In the calculation of the efficiency we use the
``right'' two charged tracks. However, if a pair of good charged tracks 
is combined with a background $\pi^0$, it does not significantly 
increase the efficiency. 

\begin{figure}
\centerline{\includegraphics[width=0.99\textwidth, height=90mm]{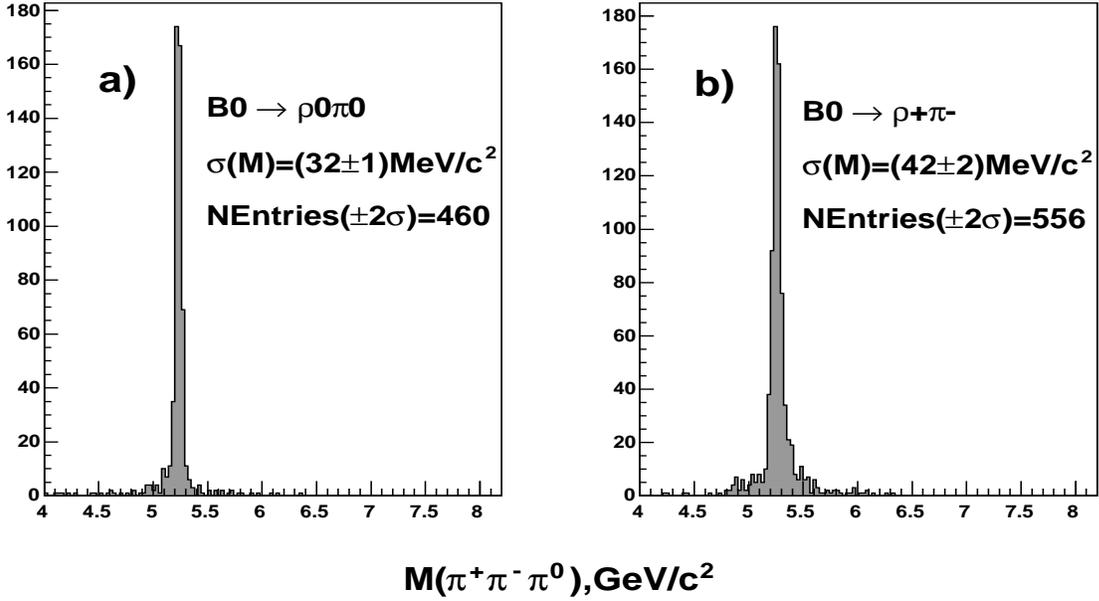}}
  \caption{Invariant mass $\pi^+ \pi^- \pi^0$ (after cuts) for the simulated 
  (a)~$B^0 \rightarrow \rho^0\pi^0$ and (b)~$B^0 \rightarrow \rho^+\pi^-$ decays.
  Each event includes on the average two Poisson distributed non-beauty interactions 
  per beam crossing, mixed to the beauty production interaction. The number of signal 
  event were counted as the number of entries in the $\pm 2 \sigma$ interval around 
  the $B$ mass, minus estimated number of background entries.}
\label{fig:brhopi}
\end{figure}

\begin{figure}
\centerline{\includegraphics[width=0.99\textwidth, height=90mm]{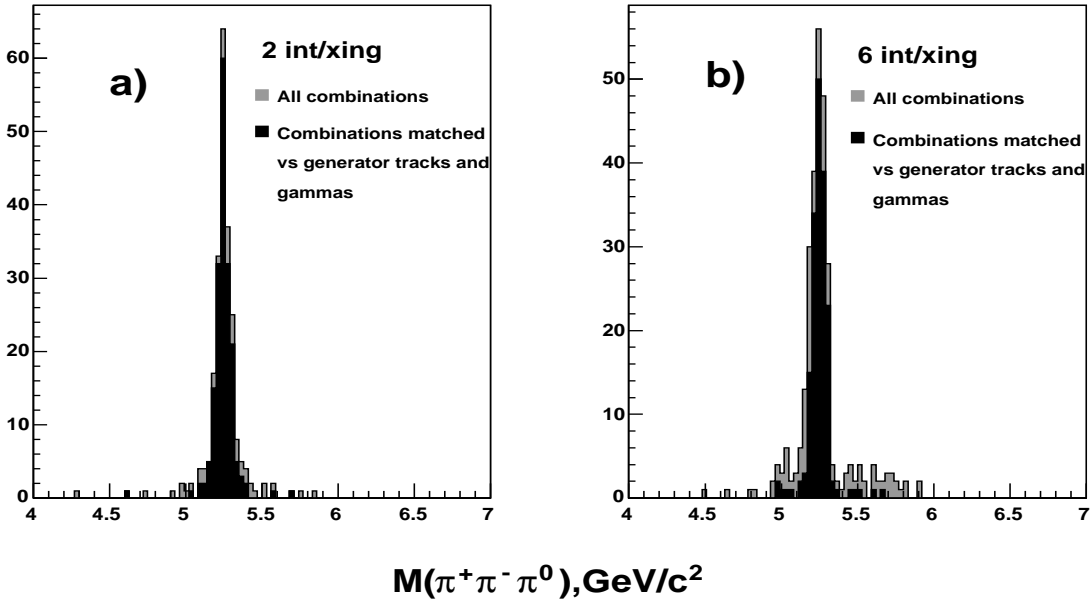}}
\caption{Invariant mass $\pi^+ \pi^- \pi^0$ (after cuts) matched vs
generator tracks and photons, for the simulated $B^0 \rightarrow \rho^+\pi^-$ 
at 2 and 6 interactions per crossing.}
\label{fig:compgen}
\end{figure}

We look for events containing a secondary vertex formed by two oppositely 
charged tracks. One of the most important selection requirements for 
discriminating the signal from the background is that the events have well 
measured primary and secondary vertices. We demand that the primary and the 
secondary vertices be well defined by requiring $\chi^2 /dof < 2$ for 
their vertex fits. 
Once the primary and the $B$ decay vertices are determined, the distance $L$
between the vertices and its error $\sigma_L$ are computed. The quantity $L/\sigma_L$
is a measure of the significance of detachement between the primary and
secondary vertices. We require $L/\sigma_L > 4$.
The two vertices must also be separated from 
each other in the plane transverse to the beam.
We define $r_{transverse}$ in terms of the primary interaction vertex position 
($x_P, y_P, z_P$) and the secondary vertex position ($x_S, y_S, z_S$), namely
$r_{transverse} = \sqrt{(x_P - x_S)^2 + (y_P - y_S)^2}$ and reject events 
when $r_transverse < 0.132$~mm 
Finaly, to insure 
that the charged tracks do not originate from the primary, we require that both
the $\pi^+$ and the $\pi^-$ candidates have an impact parameter with respect 
to the primary vertex (DCA) greater than 100~$\mu m$.

When we calculate the invariant masses of the $\pi^+ \pi^-$ and
$\pi^{\pm} \pi^0$ pairs, we require at least one of them to be compatible 
with the $\rho$ mass, that is, between 0.55 and 1.1~GeV$/c^2$.
In addition, we use several kinematic cuts which reduce the 
background to $B \rightarrow \rho \pi$ without significantly decreasing 
the reconstruction efficiency.
We require that $p_t^{\rm sum}$ divided by the scalar sum of the $p_t$ values 
of all three particles, ($p_t^{\rm sum} / \Sigma p_t$), be small.
The vector sum $p_t^{\rm sum}$ is defined with respect to the $B$ direction
of flight which is calculated from the reconstructed primary and secondary
vertices.
We also make a cut on the $B$ proper time decay requiring it to be less than
5.5 times the $B^0$ lifetime ($t_{proper} / t_0 < 5.5$).
The selection criteria are summarized in Table~\ref{tab:cuts}.


\begin{table}
\begin{center}
\caption{Selection Criteria. The notation is defined in the text. }
\label{tab:cuts}
\vspace{0.3cm}
\begin{tabular}{|l|c|c|}
\hline
Criteria                         & Value    \\
\hline
 Primary vertex criteria         & $\chi^2<2$ \\
 Secondary vertex criteria       & $\chi^2<2$ \\
 $r_{transverse}$ (cm)           & 0.0132  \\
 Normalized distance  $L/\sigma$ & $>4$ \\
 Distance $L$,~cm                & $<5$          \\
 DCA of track,~$\mu$m            & $>100$        \\
 $t_{proper}/t_0$                & $<5.5$          \\
 $E_{\pi^+},$ GeV                & $>4$          \\
 $E_{\pi^-},$ GeV                & $>4$          \\
 $p_t(\pi^+),$~GeV/$c$           & $>0.4$        \\
 $p_t(\pi^-),$~GeV/$c$           & $>0.4$        \\
 Isolation for $\gamma$,~cm      & $>5.4$        \\
 $E_{\pi^0},$ GeV                & $>5$          \\
 $p_t(\pi^0),$~GeV/$c$           & $>0.75$       \\
 $p_t^{\rm sum}/\Sigma p_t$      & $<0.06$       \\
 $m_{\gamma \gamma},$ MeV/$c^2$  & $125-145$ \\
 $m_{\pi \pi},$ GeV/$c^2$        & $0.55-1.1$\\
\hline
\end{tabular}
\end{center}
\end{table}


The results are shown in Figure~\ref{fig:brhopi} for $B^0 \rightarrow \rho^0\pi^0$ (a)
and $B^0 \rightarrow \rho^+\pi^-$ (b) Monte Carlo samples, respectively.
The $B^0$ mass resolution in these samples is in the range 32-42~MeV$/ c^2$.
The signal interval is defined as $\pm 2 \sigma$ around the $B$ mass, minus estimated
background.
The reconstruction efficiency is ($0.18 \pm 0.02$)\% for
$B^0 \rightarrow \rho^0 \pi^0$ and
($0.22 \pm 0.02$)\% for $B^0 \rightarrow \rho^{\pm} \pi^{\mp}$.

Similar simulation studies were repeated with six non-beauty interations per crossing 
mixed to the beauty production interaction to estimate 
reconstruction efficiency for the $B^0 \rightarrow \rho^+ \pi^-$ decay. Results were 
compared with those at two interactions per crossing. 
The statistics used to compare the two cases were 100,000 events.

At six interations per beam crossing the $B^0$ mass resolution remains
practicaly unchanged, as it is found to be $(44 \pm 3)$~MeV/$c^2$.  
The $B^0 \to \rho^+ \pi^-$ reconstruction efficiency is estimated at
$(0.2 \pm 0.02)$\%. This represents the effect of only 10\% as compared
to two background interactions per crossing. 

However, the number of false $3 \pi$ combinations that would pass the cuts 
appears to increase somewhat as the number of non-beauty interactions per beam 
crossing goes up. To prove that most of the entries in the $B^0$ mass
region are true $\pi^+ \pi^- \pi^0$ combinations coming from the
$B^0$ decay, we have done a check against generator level information.
Results of comparison are presented in Figure~\ref{fig:compgen}.
It is clear that the $B^0$ signal dominates in both distributions; 
false $3 \pi$ combinations could, in principal, mimic the signal but 
most of the $3 \pi$ combinations are the correct ones.

Using the previously calculated reconstruction efficiency we could expect 
to have $\sim 1000$ flavor-tagged 
$\rho^{\pm} \pi^{\mp}$  events and $\sim 150$ flavor-tagged 
$\rho^0 \pi^0$ events per year ($2 \times 10^7$~s given that BTeV was assumed
to run 10 months per year). The samples would 
include both $B^0$ and $\overline B^0$ decays, with proper time
measurements for both states.

In principal, one can use the untagged sample in the likelihood (see Section 7)
to extract~$\alpha$. Actually, this sample does not carry any information on
$\alpha$ but it allows to extract other parameters related to direct CP violation
and helps the fit converge. This leads to an improved resolution on $\alpha$ when
the untagged sample is utilized. However, in this paper we present  results obtained 
only with the tagged sample.



 
\section{Signal-to-Background Ratio in $B^0 \rightarrow (\rho\pi)^0$
decays in BTeV}

The analysis by Snyder and Quinn \cite{Snyder:1993mx} showed that with 
2,000 background free events they could always find a solution for $\alpha$.
BTeV could have collected such a statistics within 4x10$^7$ seconds 
(approximately 2 years). 
But we expect some background whose effects need to be estimated.

For a channel with a branching ratio on the order of 10$^{-5}$ and efficiencies
lower than 1\%, it is necessary to generate at least 10$^7$ {\it b\=b} 
background events. For this study we generated $2 \times 10^7$ generic 
{\it b\=b} events ($B \rightarrow \rho\pi$ channels excluded) and processed 
them through the GEANT3-based full detector simulation.
Each event contains a mean of two Poisson distributed non-beauty interactions.
Selection criteria listed in Table~\ref{tab:cuts} are applied. 
To get the background estimate, we count all of the events between 5 
and 7 GeV/$c^2$, then we scale that number down by the ratio of the
signal region divided by the background selection region.

\begin{figure}
\centerline{\includegraphics[width=0.99\textwidth,height=80mm]{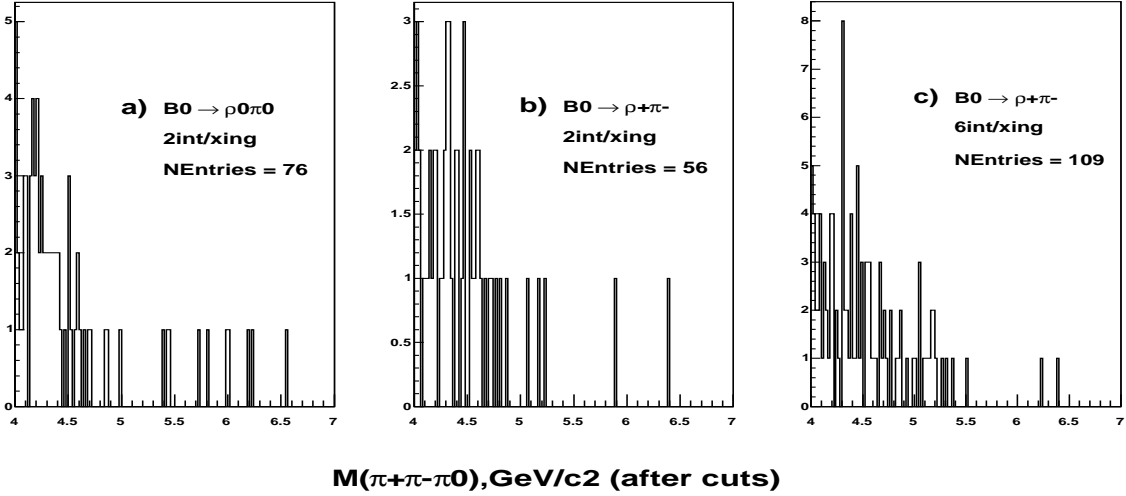}}
\caption{Background $\pi^+ \pi^- \pi^0$ invariant mass for
$B^0 \rightarrow \rho^0\pi^0$ and $B^0 \rightarrow \rho^+\pi^-$ at 2 interactions per 
crossing (a,b), and $B^0 \rightarrow \rho^+\pi^-$ at 6 interactions per crossing (c)}
\label{fig:brhopi_bck}
\end{figure}

The results of the analysis are presented in Figure~\ref{fig:brhopi_bck}a~and~b.
The signal-to-background levels are approximetely 4:1 and 1:3 for 
$\rho^{\pm}\pi^{\mp}$ and $\rho^0\pi^0$, respectively, when there are on the
average two interactions per crossing.

We have also investigated the effect of a larger number of interactions
per crossing on the $\rho^+\pi^-$ background, similar to the study on the
signal sample. We merged our background sample with an additional sample 
of non-beauty eventsgenerated with a Poisson distributed average of four
interactions per crossing. Charged tracks in the merged events were projected
onto the calorimeter, and photons from both samples were added in. Thus, the
full confusion of six interactions per crossing is simulated in the calorimeter.
The way we did this study, the confusion is not present in the simulation
of the tracking system but separate studies show that the charged particle
tracking system is reasonably robust against six interactions per crossing. 
The analysis then proceeded as before. 
We have reprocessed $1.33 \times 10^7$ events
and have compared the results corresponding to this statistics at two or six
interactions per crossing. 
The background to $B^0 \to \rho^+ \pi^-$ at six interactions per crossing is 
presented in Figure~\ref{fig:brhopi_bck}c). 
We have found that at six interactions per crossing the background to $B^0 \to \rho^+ \pi^-$
increased to 109 events, at compared to 56 events at two interactions per crossing.
This demonstrates that the background could increase at six interactions per crossing 
but the effect is expected to be about a~factor~of~2.

\section{Representation of amplitudes and phenomenological inputs}

\subsection{Classification of the amplitudes}


In this section we first define the formalism. Next, we estimate
parameter values so that the simulation is as close as possible 
to reality.

\begin{figure}[th]
\begin{center}
\epsfbox{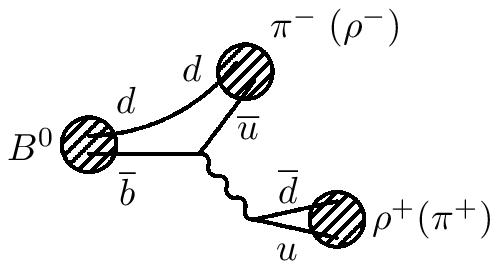} \epsfbox{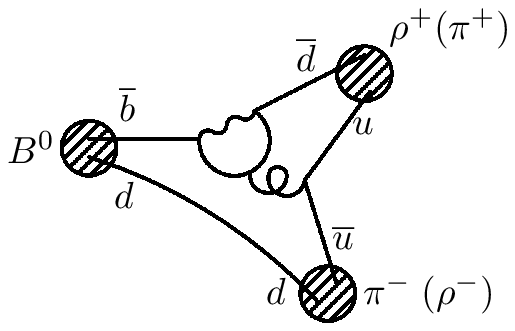} \caption{The tree (left) and
penguin (right) diagrams for the $B^0\to \rho^+\pi^-$ ($B^0\to
\rho^-\pi^+$) decays.} \label{diag}
\end{center}
\end{figure}

Amplitudes of neutral $B^0$ meson decay to $\rho \pi$ are represented 
in the form 
\begin{equation}\label{eq:amp}
    | B^0 \rangle = f_i a_{ij}, \qquad \{ij\}=\{+-\},\{-+\},\{00\},
\end{equation}
\begin{equation}
    a_{ij}= \left(-e^{-{\rm i}\alpha}T_{ij}+P_{ij}\right)e^{-{\rm
    i}\beta},
\end{equation}
where $T_{ij}$ and $P_{ij}$ give tree and penguin amplitudes,
correspondingly, as depicted in Fig. \ref{diag} extracted from
\cite{Gronau}.

The $f_k$ represents the relativistic Breit-Wigner form for $\rho \to \pi\pi$
\begin{equation}
 f_k = \frac{cos(\theta_k)}{s - m^2_\rho + i \Pi (s)},
\end{equation}
where $s$ is the square of the invariant mass ($\pi_1 , \pi_2$) and $\theta_k$
is the angle between a decay pion and the line of flight of the $\rho$.
The function $\Pi(s)$ is defined as
\begin{equation}
\Pi(s) = \frac{m^2_\rho}{\sqrt{s}} \left(\frac{p(s)}{p(m^2_\rho)}\right)^3 \Gamma(m^2_\rho), \qquad
p(s) = \sqrt{s/4 - m^2_\rho},
\end{equation}
where $m_\rho$ is the $\rho$ mass and $\Gamma$ is the width.

The amplitudes $a_{ij}$ for $B^0$ and $\overline B^0$ decay are written 
as a sum of tree ($T$) and penguin ($P$) contributions as
\begin{eqnarray}\label{eq:6amp}
 a_{+-}=-e^{i\gamma}T^{+-}+e^{-i\beta}P^{+-}  \nonumber\\
 a_{-+}=-e^{i\gamma}T^{-+}+e^{-i\beta}P^{-+} \nonumber\\
 a_{00}=-e^{i\gamma}T^{00}+e^{-i\beta}P^{00} \nonumber\\
 \bar a_{+-}=-e^{-i\gamma}T^{-+}+e^{i\beta}P^{-+} \nonumber\\
 \bar a_{-+}=-e^{-i\gamma}T^{+-}+e^{i\beta}P^{+-} \nonumber\\
 \bar a_{00}=-e^{-i\gamma}T^{00}+e^{i\beta}P^{00}, 
\end{eqnarray}
where $\gamma$ and $\beta$ are the usual CKM angles and $\alpha+\beta+\gamma =\pi$. 
Using both isospin symmetry and the fact that the Penguin amplitude is a pure 
$\Delta I = 1/2$ transition leads to the replacement 
\begin{equation}\label{eq:isop}
    P_{00}=-\frac{1}{2}\left(P_{+-}+P_{-+}\right).
\end{equation}
For the tree diagrams after Fierz transformation one naively gets
\begin{equation}\label{eq:isot}
    T_{00}=\frac{1}{2}\,\frac{a_2}{a_1}\left(T_{+-}+T_{-+}\right),
\end{equation}
where factors $a_{1,2}$ represent contributions due to gluon
corrections to the weak interactions of quarks and depend on
renormalization scale. The corresponding Lagrangian is
approximated by
\begin{equation}\label{eq:weak}
    {\mathcal L}=\frac{G_{\rm
    F}}{2\sqrt{2}}\,V_{ud}V^*_{ub}\,C_{\pm}\,(\bar b_i {\mathcal
    O}_\mu u^j)\,(\bar u_k {\mathcal O}^\mu d^l)\,\left[\delta^i_j\delta^k_l\pm
    \delta^i_l\delta^k_j\right],
\end{equation}
where indices mark SU(3)-colors, and the factors are defined as
\begin{equation}\label{eq:fact}
    a_1 = \frac{1}{2 N_c}\,[C_+ (N_c+1)+C_-(N_c-1)],\qquad
    a_2 = \frac{1}{2 N_c}\,[C_+ (N_c+1)-C_-(N_c-1)] 
\end{equation}
where $N_c=3$. In the limit of neglecting the gluon corrections we get
$$
C_{\pm}=1,\qquad a_1=1,\qquad a_2=\frac{1}{N_c}.
$$
However, the corrections taking into account the renormalization
group dependence on a decay scale put the ratio $a_2/a_1$ to a
negative value approximately given by
\begin{equation}\label{eq:ratio}
    \frac{a_2}{a_1} =-0.25\pm 0.05.
\end{equation}
In fact, as was shown in \cite{BNetc} and \cite{BenekeNeubert},  
expressions (\ref{eq:fact}) following from the factorization hypothesis can 
be significantly modified by ``non-factorizable effects'' in the complex phase. 
This means that only the absolute value of the ratio
$$
\left|\frac{a_2}{a_1}\right| =0.25\pm 0.05
$$

is reliable. The value of the phase is not realiably predicted by the theory
but is roughly estimated to be about $45^\circ$.

Corrections in (\ref{eq:isop}) and (\ref{eq:isot}) due to isospin
symmetry breaking are considered to be negligible \cite{Gronau}
and they are not included in the following simulations.

\subsection{Phenomenological constraints on the parameters}

A phenomenological analysis of possible values for the amplitudes
in (\ref{eq:amp}) has been performed in \cite{Gronau} and \cite{Charles}. 
It is based on a global fit to the measured rates assuming SU(3)-flavor 
symmetry for $B \to \rho\pi$, $B\to K^*\pi$ and $B\to \rho K$ decays. 
The analysis gives approximately twice enhancement of penguin amplitudes 
in comparison with QCD expectations \cite{BenekeNeubert}. The
preferable values of amplitudes with theoretical expectations of
uncertainties are summarized in Table~\ref{prefer}, which has
two sets of parameters we use in our modelling of the signal.

\begin{table}[th]
 \caption{Amplitudes in units of $T_{+-}$ set to $1$ and expected
 uncertainties from fits in \cite{Gronau} and \cite{Charles}.}\label{prefer}
\begin{tabular}{|c|c|c|c|}
\hline \hspace*{5mm}Parameter\hspace*{5mm} & \hspace*{5mm}Set
I\hspace*{5mm} & \hspace*{5mm}Set II\hspace*{5mm} &
Theor. uncertainties or limits\\
\hline &&&\\[-2mm]
$\quad |T_{-+}|$  & 0.8 & 0.8 & $0.63-0.9$ \\[2mm]
$\arg\left[T_{-+}\right]$ & $-20^\circ$ & $-20^\circ$ & $\pm 10^\circ$\\[2mm]
$\quad |P_{+-}|$ & 0.18 & 0.18 & $\pm 0.05$ \\[2mm]
$\arg[P_{+-}]$ & $30^\circ$ & $30^\circ$ & $\pm 30^\circ$ \\[2mm]
$\quad \displaystyle\left|\frac{P_{-+}}{T_{-+}}\right|$ & 0.28 & 0.28 &
$0.14-0.32$\\[4mm]
$\displaystyle\arg\left[\frac{P_{-+}}{T_{-+}}\right]$ &
$80^\circ$ & $130^\circ$ & $\pm 60^\circ$\\[3mm]
$\quad \displaystyle\left|\frac{a_2}{a_1}\right|$ & 0.25 & 0.25 & $0.18-0.32$\\[3mm]
$\displaystyle\arg\left[\frac{a_2}{a_1}\right]$ & 45$^\circ$ & 45$^\circ$ & $0-2\pi$\\[3mm]
$\alpha$ & $88^\circ$ & $100^\circ$ & $80^\circ-110^\circ$ \\[1mm]
 \hline
\end{tabular}
\end{table}

The value of $|T_{-+}|$ is ordinary fixed by the factorization
hypothesis \cite{Gronau}, so that it is equal to the ratio of
decay constants $f_\pi/f_\rho\approx 0.63$, while the ratio fitted 
by SU(3) ansatz results in a greater value of about $0.7$. 
Nevertheless, we fix this parameter to
$0.8$, reproducing the mean magnitude of branching fractions. The
uncertainties of complex phases 
are not given explicitly in \cite{Gronau} and \cite{Charles}, but we expect
them to be lower than $30^\circ$ at fixed absolute
values of penguin-to-tree ratios. The amplitude of
$B^0\to\rho^0\pi^0$ is constructed in accordance with
(\ref{eq:isop}), (\ref{eq:isot}) and (\ref{eq:ratio}).

Now we make the transition to the estimate of parameter values 
in the simulation and to the comparison with the existing experimental 
results.

The overall normalization is tuned to the experimental sum of
CP-averaged branching ratios
$$
{\mathcal B}^{\pm\mp}_{\rho\pi}=
{\mathcal B}^{+-}_{\rho\pi}+{\mathcal B}^{-+}_{\rho\pi}=(24.0\pm2.5)\times 10^{-6},
$$
where
$$
{\mathcal B}^{+-}_{\rho\pi}=\frac{1}{2}\left\{{\mathcal B}[B^0\to
\rho^+\pi^-]+{\mathcal B}[\bar B^0\to
\rho^-\pi^+]\right\}=(13.9\pm2.2)\times 10^{-6},
$$
$$
{\mathcal B}^{-+}_{\rho\pi}=\frac{1}{2}\left\{{\mathcal B}[B^0\to
\rho^-\pi^+]+{\mathcal B}[\bar B^0\to
\rho^+\pi^-]\right\}=(10.1\pm2.1)\times 10^{-6}.
$$
For example, taking the set I of parameters we get
$$
{\mathcal B}[B^0\to \rho^+\pi^-]=16.5\times 10^{-6},\qquad
 {\mathcal B}[B^0\to \rho^-\pi^+]=14.1\times 10^{-6},\qquad
 {\mathcal B}[B^0\to \rho^0\pi^0]=0.9\times 10^{-6},
$$
and
$$
{\mathcal B}[\bar B^0\to \rho^+\pi^-]=4.6\times 10^{-6},\qquad
 {\mathcal B}[\bar B^0\to \rho^-\pi^+]=11.6\times 10^{-6},\qquad
 {\mathcal B}[\bar B^0\to \rho^0\pi^0]=1.4\times 10^{-6},
$$
which should be compared with experimental averages from BELLE,
BABAR and CLEO in \cite{Gronau} and \cite{Charles}
$$
{\mathcal B}[B^0\to \rho^+\pi^-]=(16.5^{+3.1}_{-2.8})\times
10^{-6},\qquad
 {\mathcal B}[B^0\to \rho^-\pi^+]=(15.4^{+3.2}_{-2.9})\times 10^{-6},\qquad
$$
and
$$
{\mathcal B}[B^0\to \rho^+\pi^-]=(4.8^{+2.6}_{-2.3})\times
10^{-6},\qquad
 {\mathcal B}[B^0\to \rho^-\pi^+]=(11.4^{+2.8}_{-2.6})\times 10^{-6}.\qquad
$$
The above branching ratios give CP-averaged values of
$$
{\mathcal B}^{+-}_{\rho\pi}=14.0\times 10^{-6},\qquad {\mathcal
B}^{-+}_{\rho\pi}=9.3\times 10^{-6}
$$
for the charged modes, while for the neutral mode we have
$$
{\mathcal B}^{00}_{\rho\pi}=\frac{1}{2}\left\{{\mathcal B}[B^0\to
\rho^0\pi^0]+{\mathcal B}[\bar B^0\to \rho^0\pi^0]\right\}=1.2\times
10^{-6}
$$
are consistent with the experimental value
$$
{\mathcal B}^{00}_{\rho\pi}=(1.7\pm 0.8)\times 10^{-6} < 2.5\times
10^{-6}\;\mbox{at 95\% C.L.}
$$
Note, that the neutral CP-averaged mode weakly depends on the
complex phase of $a_2/a_1$, but branching ratios of $B^0$ and
$\bar B^0$ strongly depend on that phase: for instance, putting
$\arg[a_2/a_1]=\pi$ gives ${\mathcal B}[B^0\to \rho^0\pi^0]=2.1\times
10^{-6}$ and ${\mathcal B}[\bar B^0\to \rho^0\pi^0]=0.2\times 10^{-6}$.


The time-dependent CP asymmetry is given by
\begin{eqnarray}\label{eq:cp_time}
a^{\pm}_{CP} & = & \frac{\Gamma(\overline B^0(t) \to \rho^{\pm}
\pi^{\mp}) - \Gamma(B^0(t) \to \rho^{\pm}\pi^{\mp})}
{\Gamma(\overline B^0(t) \to \rho^{\pm} \pi^{\mp}) + \Gamma(B^0(t) \to \rho^{\pm}\pi^{\mp})} = \nonumber\\
& & = (S_{\rho\pi} \pm \Delta S_{\rho\pi}) \sin(\Delta m_d t) -
(C_{\rho\pi} \pm \Delta C_{\rho\pi}) \cos(\Delta m_d t).
\end{eqnarray}
In this formula $S_{\rho\pi}$ and $C_{\rho\pi}$ represent
mixing-induced CP violation and flavor-dependent direct CP
violation, respectively. The value of $\Delta S_{\rho\pi}$ and $\Delta
C_{\rho\pi}$ are CP-conserving. The $\Delta C_{\rho\pi}$
characterizes the asymmetry between rates $\Gamma (B^0 \to
\rho^+\pi^-) + \Gamma (\overline B^0 \to \rho^-\pi^+)$ and $\Gamma
(B^0 \to \rho^-\pi^+) + \Gamma (\overline B^0 \to \rho^+\pi^-)$ at
$t=0$, i.e. at initial moment of evolution, while $\Delta
S_{\rho\pi}$ indicates mixing of decays at $t\neq 0$, and as we
have found, it strongly depends on both the relative strong
phase of penguin with respect to tree amplitude in $\rho^-\pi^+$
mode (the parameter $\mbox{arg}[P_{-+}/T_{-+}]$) and the CKM angle
$\alpha$.

Time-integrated asymmetries are given by
\begin{equation}\label{eq:cp_a_plmn}
A^{+-}_{\rho\pi} = - \frac{A_{\rho\pi} + C_{\rho\pi} + A_{\rho\pi} \Delta C_{\rho\pi}}
{1 + \Delta C_{\rho\pi} + A_{\rho\pi} C_{\rho\pi}} =
\frac{N(\overline B^0 \to \rho^- \pi^+) - N(B^0 \to \rho^+ \pi^-)}
{N(\overline B^0 \to \rho^- \pi^+) + N(B^0 \to \rho^+ \pi^-)},
\end{equation}
\begin{equation}\label{eq:cp_a_mnpl}
A^{-+}_{\rho\pi} = - \frac{A_{\rho\pi} - C_{\rho\pi} - A_{\rho\pi} \Delta C_{\rho\pi}}
{1 - \Delta C_{\rho\pi} - A_{\rho\pi} C_{\rho\pi}} =
\frac{N(\overline B^0 \to \rho^+ \pi^-) - N(B^0 \to \rho^- \pi^+)}
{N(\overline B^0 \to \rho^+ \pi^-) + N(B^0 \to \rho^- \pi^+)} ,
\end{equation}
\begin{equation}\label{eq:cp_a_00}
A_{\rho\pi} = \frac{ |a_{+-}|^2 + |\bar a_{+-}|^2 - |a_{-+}|^2 -
|\bar a_{-+}|^2 } { |a_{+-}|^2 + |\bar a_{+-}|^2 + |a_{-+}|^2 +
|\bar a_{-+}|^2 }      ,
\end{equation}
where non-zero values of $A^{+-}_{\rho\pi}$ and $A^{-+}_{\rho\pi}$ indicate direct
CP violation. 

In Table~\ref{refer2} we show a comparison between observable quantities 
obtained by using sets I and II values with experimental data.

\begin{table}[th]
  \centering
  \caption{A comparison of values of quantities evaluated from sets I and II
  with available experimental data from \cite{Charles}.}\label{refer2}
    \begin{tabular}{|c|c|c|c|}
      \hline
      Quantity, units & Set I & Set II & Exp.data \\
      \hline
      ${\mathcal B}[B^0\to \rho^+\pi^-],\;10^{-6}$ & 16.5 & 16.5 &
      $16.5^{+3.1}_{-2.8} $\\[2mm]
      ${\mathcal B}[B^0\to \rho^-\pi^+],\;10^{-6}$ & 14.1 & 14.3 &
      $15.0^{+3.2}_{-2.9} $\\[2mm]
      ${\mathcal B}[B^0\to \rho^0\pi^0],\;10^{-6}$ & 0.9 & 0.6 &
      $1.7\pm0.8\, (\mathcal{B}^{00}_{\rho\pi})$\\[2mm]
      ${\mathcal B}[\bar B^0\to \rho^+\pi^-],\;10^{-6}$ & 4.6 & 6.6 &
      $4.8^{+2.6}_{-2.3} $\\[2mm]
      ${\mathcal B}[\bar B^0\to \rho^-\pi^+],\;10^{-6}$ & 11.6 & 11.4 &
      $11.6^{+2.8}_{-2.6} $\\[2mm]
      ${\mathcal B}[\bar B^0\to \rho^0\pi^0],\;10^{-6}$ & 1.4 & 1.8 &
      $1.7\pm 0.8\, (\mathcal{B}^{00}_{\rho\pi}) $\\[2mm]
      $A_{\rho\pi}$ & -0.100 & -0.054 & $-0.114\pm 0.067$ \\[2mm]
      $S_{\rho\pi}$ & -0.15 & -0.30 & $-0.13\pm 0.18$ \\[2mm]
      $\Delta S_{\rho\pi}$ & 0.33 & 0.35 & $0.33\pm 0.18$ \\[2mm]
      $C_{\rho\pi}$ & 0.33 & 0.27 & $0.35\pm 0.14$ \\[2mm]
      $\Delta C_{\rho\pi}$ & 0.24 & 0.16 & $0.20\pm 0.14$ \\[2mm]
      $A_{\rho\pi}^{+-}$ & -0.17 & -0.18 & $-0.18\pm 0.14$ \\[2mm]
      $A_{\rho\pi}^{-+}$ & -0.51 & -0.37 & $-0.52^{+0.18}_{-0.20}$ \\[2mm]
      $\alpha$ & $88^\circ$ & $100^\circ$ &
      $100^{\circ+12^\circ}_{~-10^\circ}$,\\ &&& (\mbox{CKM
      unitarity:} $\,98^\circ\pm 16^\circ$)\\
      \hline
    \end{tabular}
\end{table}

The comparison shows that the range of parameters we use seem to be reasonable.



\section{Time-dependent Dalitz Plot Analysis of $B^0 \rightarrow (\rho\pi)^0$ 
decays in BTeV}

In section 4 we demonstrated that BTeV would be able to collect
a sample of $\sim1000$ flavor-tagged $B^0 \to (\rho\pi)^0$ 
events within one year of operation, which would allow a reliable Dalitz 
plot analysis of this decay mode.


The model of the Dalitz plot analysis has three parts :
\begin{itemize}
\item{ $B \to \rho \pi$ signal}
\item{ random true $\rho$ plus random true $\pi$, the ``resonant background''}
\item{ uniform density, the ``non-resonant background''}
\end{itemize}

The formalism used to fit the Dalitz is based on 13 independent parameters : 
6 amplitudes, 6 strong phases and the weak phase $\alpha$ itself. 
Using the constrains given in eq.~\ref{eq:6amp} and \ref{eq:isop} we can 
reduce the number of parameters to 11. 
We fix a reference rate and strong phase so that the total number of
parameters reduces to 9. Two additional parameters must be added if we 
allow the resonant and non-resonant background fractions to be determined
by the fit.

Due to the low reconstruction efficiency of this particular final state 
it would not be feasible to include the Snyder-Quinn formalism directly
into the full detector Monte Carlo simulation: it would have required 
significant computer power and the generation of a huge number of events
to obtain the desired statistics.

We have opted for a different approach. 
The generated template events are distributed flat over the Dalitz plot 
domain, with the exponential time distribution and random tag=$\pm$1.
We further use a rejection algorithm based on the isospin amplitudes
formalism for the tree and the penguin contributions to the Dalitz plot.

Time evolution of the $B^0 \to (\rho \pi)^0$ decay
amplitudes, including $B- \overline B$-mixing, is given~by~:
\begin{equation}
\mathcal{A} = e^{- \Gamma t / 2} ( \cos {{\Delta M t}\over{2}} | B^0 \rangle 
+ i {{q}\over{p}} \sin {{\Delta M t}\over{2}} | \overline{B}^0 \rangle )
\end{equation}
\begin{equation}
\overline{\mathcal{A}} = e^{- \Gamma t / 2} ( i {{p}\over{q}} \sin {{\Delta M t}\over{2}} | B^0 \rangle 
+ \cos {{\Delta M t}\over{2}} | \overline{B}^0 \rangle )
\end{equation}
where $| B^0 \rangle$ is given by eq.\ref{eq:amp}.
The template events are accepted or rejected based on whether a random number
is less than or greater than $|\mathcal{A}|^2 / |\mathcal{A}_{max}|^2$.

The background has been parameterized to account for both non-resonant and resonant
components. The non-resonant background has been uniformly distributed over the Dalitz 
plot domain. The resonant background allows the two pions to have a Breit-Wigner shaped
enhancement with the $\rho$ line shape.

The process of reconstruction of the accepted events is simulated by smearing
them using the resolutions on momentum reconstruction and lifetime. These 
values were obtained from the simulation described in Section 4. The smearing 
has been computed comparing the reconstructed momentum, of $\pi^+$, $\pi^-$ 
and $\pi^0$ to the generator information, $\sigma (p_{gen} - p_{rec})/p_{gen}$ : 
0.7\% for charged pions and at 0.9\% for $\pi^0$'s. 
Signal events are generated with an exponential time distribution.
The rejection algorithm appropriately shapes the time evolution of the $B^0$'s 
according to mixing and CP violation.
The resolution on lifetime has also been estimated to be 64~fs using the Monte Carlo 
described in section 4, by computing 
the reconstructed lifetime and comparing it to the generated one.  
It should be pointed out that the resolution on lifetime
is independent of lifetime. Proper time dependent acceptance was included in 
the likelihood. 
The background level is determined by a full Geant simulation of
20,000,000 generic $b \overline b$ events; it is assumed that this background 
has an exponential time dependence given by the average lifetime of $b$-flavored
hadrons. 



\begin{figure}
\centerline{\includegraphics[width=0.99\textwidth,height=160mm]{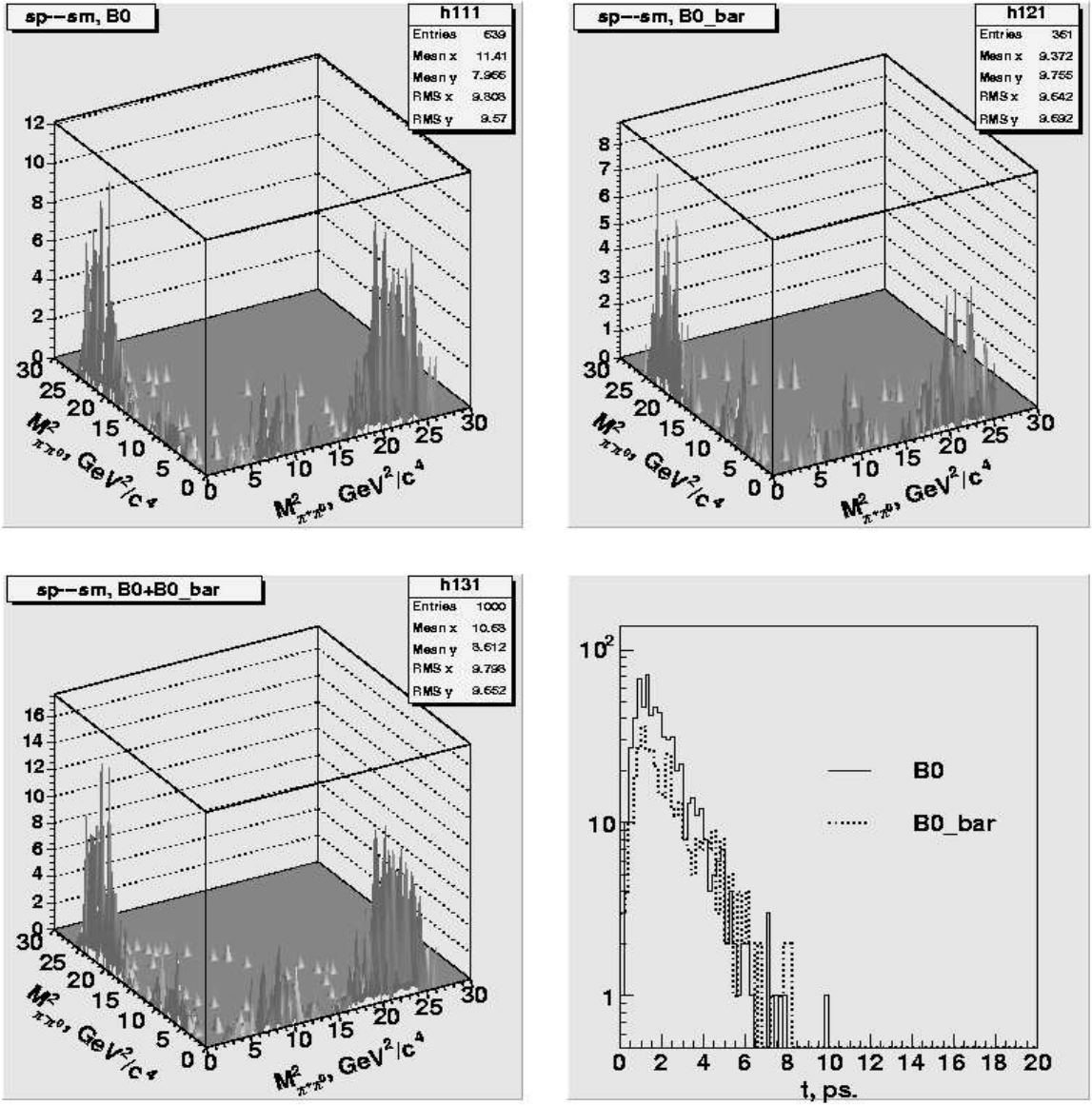}}
  \caption{Dalitz-plot and proper time distribution for one trial of 1000 events 
  for $\alpha = 88^{\circ}$ (detector efficiency included).}
\label{fig:dalitz_sample}
\end{figure}

We have used two values for $\alpha$ : $\alpha = 88^{\circ}$ and 
$\alpha = 100^{\circ}$. For each case we have generated 500 independent trials,
starting with different random numbers every time.
Every trial contains 1000 signal events, 250 non-resonant background events and 
250 resonant background events. This corresponds to one year ($2 \times 10^7$s) 
of data taking. 
The background level was chosen based on the MonteCarlo studies described
in section 5, for the case of of running at 396 ns bunch spacing (6 interactions 
per beam crossing), which would be BTeV's most challenging scenario.
The Dalitz plot for one such sample is shown in Fig.~\ref{fig:dalitz_sample}. 

To extract the parameters and the associated errors we have used an unbinned 
maximum likelihood fit. The likelihood over the full Dalitz domain is given by
\begin{equation}\label{eq:likelihood}
-2ln \mathcal L = -2 \sum_{i=1}^{N_{B^0_d}} ln \frac{\mathcal F_i}{1+R_{non}+R_{res}}
- 2 \sum_{j=1}^{N_{\overline B^0_d}} ln \frac{\overline {\mathcal F_j}}{1+R_{non}+R_{res}}
\end{equation}
where
\begin{eqnarray}
\mathcal F_i & = & 
\frac{|\mathcal A (s^+_i,s^-_i,t_i,\alpha ,..)|^2}{\mathcal N (\alpha ,..)} 
\times \varepsilon (s^+_i,s^-_i) +  \nonumber\\
& & + \left[ R_{non} \times \frac{1}{\mathcal N_t} +
R_{res} \times \frac{|{\mathcal BW}(s^+_i,s^-_i)|^2}{\mathcal N_{BW}} \right]
\times \varepsilon (s^+_i,s^-_i)
\end{eqnarray}
\begin{eqnarray}
\overline {\mathcal F}_j & = & 
\frac{| \overline{\mathcal A}(s^+_j,s^-_j,t_j,\alpha ,..)|^2}{\mathcal N (\alpha ,..)} 
\times \varepsilon (s^+_j,s^-_j) +  \nonumber\\
& & + \left[ R_{non} \times \frac{1}{\mathcal N_t} +
R_{res} \times \frac{|{\mathcal BW}(s^+_j,s^-_j)|^2}{\mathcal N_{BW}} \right]
\times \varepsilon (s^+_j,s^-_j)
\end{eqnarray}
Here $s^+_j = (m_{\pi^+} + m_{\pi^0})^2_j$ and $s^-_j = (m_{\pi^-} + m_{\pi^0})^2_j$ 
are two Dalitz plot variables for the $j$-th event.
$N_{B^0_d}$ and $N_{\overline B^0_d}$ are the total number of the $B^0_d$ and
$\overline B^0_d$ events. The normalization 
${\mathcal N} = (| {\mathcal A}|^2 + | \overline {\mathcal A} |^2) \times \varepsilon$
is integrated over the Dalitz plane and over the proper time and weighted by the 
detector efficiency. $R_{non}$ and $R_{res}$ are defined as
\begin{equation}
R_{non} = \frac {N^{back}_{non}}{N^{signal}_{B^0_d + \overline B^0_d}} \qquad
R_{res} = \frac {N^{back}_{res}}{N^{signal}_{B^0_d + \overline B^0_d}}
\end{equation}

\begin{figure}
\centering
  \subfigure[Input value $\alpha = 88^{\circ}$]{\includegraphics[width=0.95\textwidth,height=65mm]{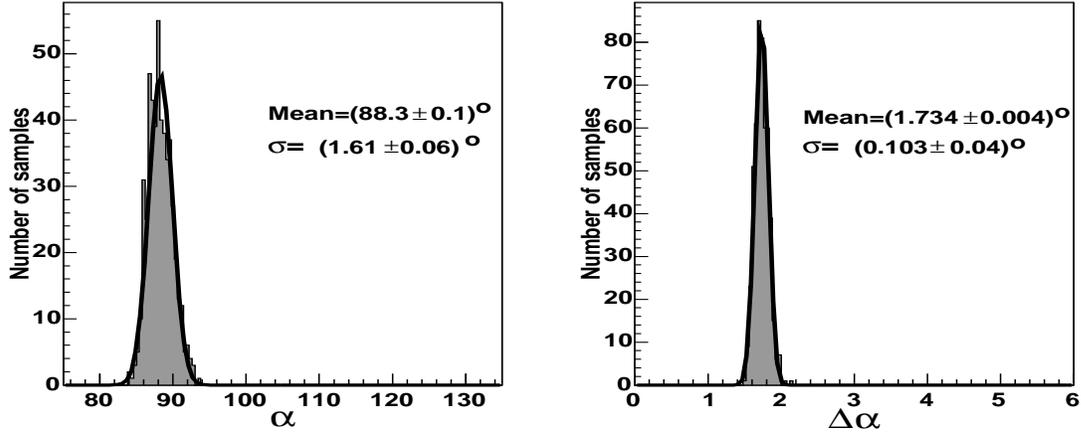}} \\
  \subfigure[Input value $\alpha = 100^{\circ}$]{\includegraphics[width=0.95\textwidth,height=65mm]{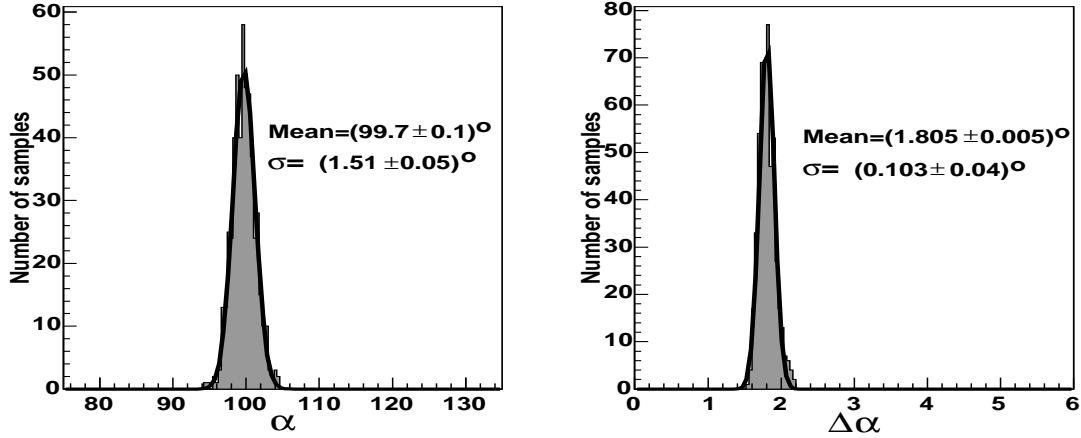}}
  \caption{Accuracy in determining $\alpha$ : fit value of $\alpha$ and deviation from input 
           value $(\alpha_{input} - \alpha_{fit})$}.
  \label{fig:alpha}
\end{figure}

The fit has been performed
for 500 samples, for each input value of $\alpha = 88^{\circ}$ and $\alpha = 100^{\circ}$, 
to confirm the stability of the procedure. 
The results of the fit value for $\alpha$ (right) and its deviation from the 
input value (left) are presented in Fig.~\ref{fig:alpha}, for the input
values of $\alpha = 88^{\circ}$ (Sample 1, Fig.~\ref{fig:alpha}a) and $\alpha = 100^{\circ}$ 
(Sample 2, Fig.~\ref{fig:alpha}b), respectively. 
We measured $\alpha = (88.3 \pm 1.6)^{\circ}$ and $\Delta \alpha = (1.7 \pm 0.09)^{\circ}$ 
for Sample 1 and $\alpha = (99.7 \pm 1.5)^{\circ}$ and $\Delta \alpha = (1.8 \pm 0.1)^{\circ}$ 
for Sample 2, which is in good agreement with the input parameters. 

These results have been obtained for ideal tagging. We have also made fits
for different errors in the tagging dilution factor. The results are presented in 
Table~\ref{tab:dilution} in the range from almost  ideal $\sigma_{dil} = 2$\% 
to the very conservative $\sigma_{dil} = 25$\%. The displacements can be considered 
as systematic errors and should be summed with the ideal tagging error in quadrature. 
Assuming that the error in the tagging dilution factor in BTeV would have been 10-15\%, 
we estimate the accuracy in measuring $\alpha$ at $(1.8-2.3)^{\circ}$ for $\alpha = 88^{\circ}$ 
and $(3.4-4.7)^{\circ}$ for $\alpha = 100^{\circ}$.
We consider the most conservative case, a 15\% error on the dilution factor for 
$\alpha = 100^{\circ}$, and conclude that BTeV could measure $\alpha$ with the 
accuracy of better that $5^{\circ}$ in one year of operation.

\begin{table}
\begin{center}
\caption{The change (in degrees) of the value of the $\alpha$
coming out of the likelihood fit at different errors in the tagging
dilution factor (in \%).}
\label{tab:dilution} 
\begin{tabular}{|l|c|c|c|c|c|c|}
\hline
& \multicolumn{6}{|c|}{$\sigma_{dil}$,\%} \\ \cline{2-7}
input $\alpha$, degrees &  2   & 5    & 10   & 15   & 20   & 25   \\ \hline 
& \multicolumn{6}{|c|}{additional displacement, degrees} \\ \cline{2-7}
88  & +1.0 & +0.4 & -0.6 & -1.6 & -2.7 & -3.9 \\ \hline
100 & -0.9 & -1.6 & -2.9 & -4.3 & -5.8 & -7.2 \\ \hline
\end{tabular}
\end{center}
\end{table}

\begin{figure}
\centerline{\includegraphics[width=0.99\textwidth]{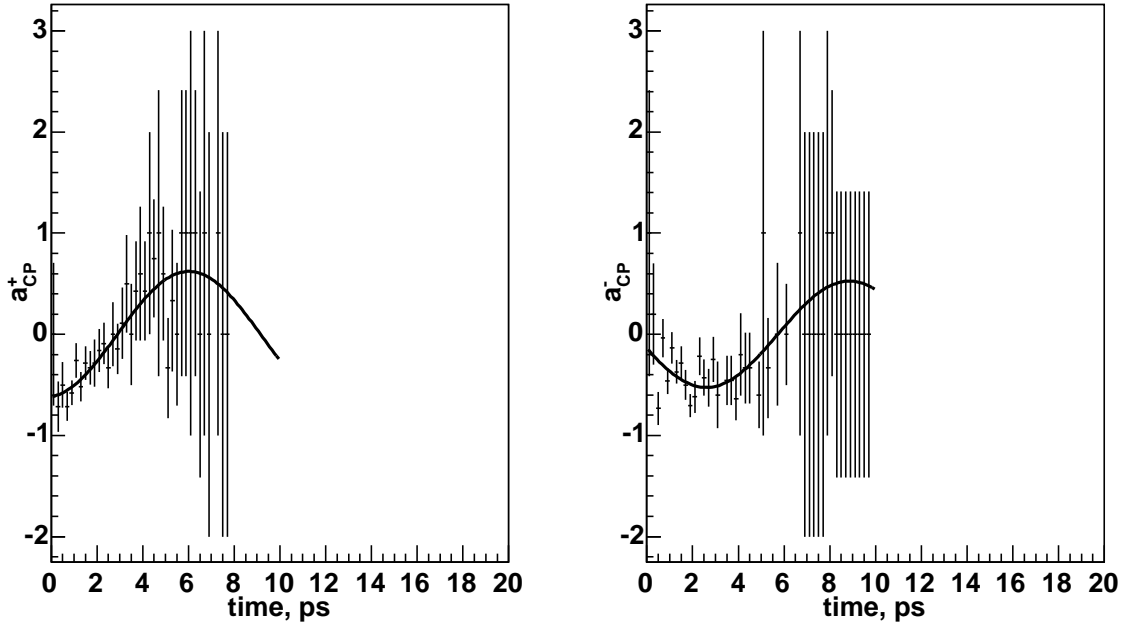}}
  \caption{CP-asymmetries $a_{cp}^+$ for $\rho^+\pi^-$ and $a_{cp}^-$ 
           for $\rho^-\pi^+$. The solid curves represent the fit.}
\label{fig:acp}
\end{figure}

In order to ensure that our event generation model is correct 
(see eq.~\ref{eq:cp_time}-\ref{eq:cp_a_00}) we have integrated
eq.~\ref{eq:likelihood} over the Dalitz-plot
domain, so that only the proper time dependence of the $B$-meson decay
rate is left. We fit the data with the assumption of the time-dependent 
CP asymmetry of the $B$-meson decay as expressed in eq.~\ref{eq:ratio}. 
The CP asymmetries $a_{cp}^+$ for $\rho^+\pi^-$ and $a_{cp}^-$ for 
$\rho^-\pi^+$ (see eq.~\ref{eq:cp_time}) are shown in Fig.~\ref{fig:acp} :
the solid curves represent the fit results. The results obtained for the
parameters in eq.~\ref{eq:cp_time} are :
\begin{equation}
S_{\rho\pi} = -0.22 \pm 0.06 (-0.15)
\end{equation}
\begin{equation}
\Delta S_{\rho\pi} = 0.29 \pm 0.06 (0.33)
\end{equation}
\begin{equation}
C_{\rho\pi} = 0.38 \pm 0.05 (0.34)
\end{equation}
\begin{equation}
\Delta C_{\rho\pi} = 0.24 \pm 0.05 (0.23)
\end{equation}
where the numbers in parentheses represent the input to the simulation.

The results of the fit are in good agreement with the input values
used in the Monte Carlo simulation. This justifies the validity of the 
model we used 
to extract $\alpha$ and demonstrates that these important parameters 
could have been measured with the accuracy of 0.05-0.06 in one year 
of BTeV operation. As one can see in Table~\ref{refer2}, up until now 
these values are known only crudely.

\section{Conclusions}

Physics simulations of the decay $B \to \rho\pi$ for the BTeV project
at Fermilab has been performed. The main idea was to estimate the expected
accuracy in extracting the angle $\alpha$ of the UT.

To calculate the signal-to-background ratio for the decay of interest,
$2 \times 10^7$ background events were simulated and processed through the
full detector simulation based on Geant3 package. 
Using the excellent electromagnetic 
calorimeter based on Lead Tungstate crystals, the $B^o \to \rho^+\pi^-$
decay the signal-to-background ratio is estimate at 4:1 or 2:1, for 132 ns 
or 396 ns beam crossing intervals, respectively.

A phenomenological analysis has been made for the possible values of 
tree and penguin amplitudes and phases for the process of interest,
based on a global fit with SU(3)-flavor asymmetry for $B \to \rho\pi$,
$B \to K^* \pi$ and $B \to \rho K$. The latest experimental data from
BaBar and BELLE were used in this analysis.

Dalitz-plot analysis of the $B \to \rho\pi$ decay with input from
the phenomenological analysis has been presented. It has been shown
that in one year  of data taking BTeV could achieve the accuracy better 
than $5^{\circ}$ on the angle $\alpha$.

The interference between tree and penguin diagrams can be exploited
by measuring the time dependent CP violating effects in the $B \to \rho\pi$
decays. In this paper it has been found that mixing-induced CP violation
parameter $S_{\rho\pi}$ and flavor-dependent direct CP violating
parameter $C_{\rho\pi}$ could be measured with the accuracy of 0.05-0.06.

\section{Acknowledgments.}   
This work was partially supported by the U.S. National Science  
Foundation and the Department of Energy
and the Russian Foundation for Basic Research grant 02-02-39008. 
We thank Alexander Bondar, Olivier Deschamps and Daniele Pedrini
for fruitfull discussions.


\end{document}